\documentclass[12pt, oneside]{article}
\usepackage{yfonts}
\usepackage{amsmath,amssymb,amsthm}
\usepackage{units,stmaryrd}
\usepackage{graphicx}
\usepackage[all]{xy}

\title{On the necessity of complexity}
\author{Joost J. Joosten\\
Department of Logic, History and Philosophy of Science\\
Faculty of Philosophy\\
Carrer Montalegre 6, 08001 Barcelona, Spain\\
\texttt{jjoosten@ub.edu}}

\begin{document}

\maketitle

\begin{abstract}
Wolfram's Principle of Computational Equivalence \index{Principle of Computational Equivalence} ($\mathbf{PCE}$) implies that universal complexity abounds in nature. This paper comprises three sections. In the first section we consider the question \textit{why} there are so many universal phenomena around. So, in a sense, we seek a driving force behind the $\mathbf{PCE}$ if any. We postulate a principle $\mathbf{GNS}$ that we call the \textit{Generalized Natural Selection} principle that together with the Church-Turing thesis is seen to be equivalent in a sense to a weak version of $\mathbf{PCE}$.
\index{Church-Turing thesis}
\index{mathematical logic}
\index{Wolfram, S.}
In the second section we ask the question why we do not observe any phenomena that are complex but not-universal. We choose a cognitive setting to embark on this question and make some analogies with formal logic.

In the third and final section we report on a case study where we see rich structures arise everywhere.
\end{abstract}


\section{Why complexity abounds}
Throughout the literature one can find various different and sometimes contradicting definitions of what complexity is. The definition that we shall employ in the first section involves the notion of a universal computational process/device. For the second and third section of this paper we shall use slightly less formal and more relaxed notions of the word complexity.

\subsection{What is complexity?}
Let us recall that a computational process $\Pi$ is universal \index{universal computational process} if it can simulate any other computational process $\Theta$. In other words, $\Pi$ is universal if for any other computational process $\Theta$, we can find an easy coding protocol $\mathcal C$ and decoding protocol $\mathcal C^{-1}$ so that we can encode any input $x$ for $\Theta$ as an input $\mathcal C(x)$ for $\Pi$ so that after $\Pi$ has performed its computation we can decode the answer $\Pi ({\mathcal{C}(x)})$ to the answer that $\Theta$ would have given us. In symbols: $\mathcal{C}^{-1}(\Pi (C(x)))=\Theta(x)$.

For the sake of this first section we can take as working definition that a system is complex if we can easily perceive it as a universal computational process. Note that we have used the word `easy'  a couple of times above. If we were to be more precise we should specify this and could for example choose for \textit{poly-time} or some other technical notion that more or less covers the intuition of what is easy. We wish to be not too specific on these kind of details here.

Thus, for the first section of this paper, a complex process is one that is computationally universal.  However, great parts of the reasoning here will also hold for other definitions of complexity. For example, stating that a process is complex if comprehending or describing it exceeds or supersedes all available resources (time, space, description size).

Note that our current definition of complexity need not necessarily manifest itself in a complex way. Remember that a universal process is one that can simulate \textit{any} other process, thus also including the very simple and repetitive ones. One might thus equally well observe a universal process that temporarily exhibits very regular behavior. In this sense universal cannot be directly equated to our intuitive notion of complexity but rather to \textit{potentially complex}.

\subsection{The Principle of Computational Equivalence and the Church-Turing thesis}

In his NKS book \cite{NKS}, Wolfram postulates the Principle of Computational Equivalence ($\mathbf{PCE}$):

\begin{quote}
{\bf $\mathbf{PCE}$}: Almost all processes that are not obviously simple can be viewed as computations of equivalent and maximal sophistication. \index{Principle of Computational Equivalence}
\end{quote}

The processes here referred to are processes that occur in nature, or at least, processes that could in principle be implemented in nature. Thus, processes that require some oracle or black box that give the correct answer to some hard questions are of course not allowed here.

As noted in the book, $\mathbf{PCE}$ implies the famous Church-Turing Thesis \index{Church-Turing Thesis} ($\mathbf{CT}$):

\begin{quote}
{\bf $\mathbf{CT}$}: Everything that is algorithmically computable is computable by a Turing Machine.\index{Turing machine}\index{universal computation}
\end{quote}


In Section \ref{section:TMs} below we shall briefly revisit the definition of a Turing Machine. Both theses have some inherent vagueness in that they try to capture/define an intuitive notion. While the $\mathbf{CT}$ thesis aims at defining the intuitive notion of algorithmic computability, $\mathbf{PCE}$ aims at defining what degrees of complexity occur in natural processes. But note, this is not a mere definition as, for example, the notion of what is algorithmically computable comes with a clear intuitive meaning. And thus, the thesis applies to all such systems that fall under our intuitive meaning. 
\index{DNA computing}
As a consequence, the $\mathbf{CT}$ thesis would become false if some scientists were to point out an algorithmic computation that cannot be performed on a Turing Machine with unlimited time and space resources. With the development and progress of scientific discovery the thesis has to be questioned and tested time and again. And this is actually what we have seen over the past decades with the invention and systematic study of new computational paradigms like DNA computing \cite{DNAcomputing}, quantum computing \cite{QuantumComputing}, membrane computing \cite{MembraneComputing},  etc. Most scientists still adhere to the $\mathbf{CT}$ thesis. There are some highly theoretical notions of super-computations and super-tasks which would allegedly escape the $\mathbf{CT}$ thesis but to my modest esteem, they depend too much on very strong assumptions and seem impossible to be implemented \cite{Hypertasks}. However, I would love to be proven wrong in this and see such a super-computer be implemented.

In the $\mathbf{PCE}$ there is moreover a vague quantification present in that the principle speaks of \textit{almost all}. This vague quantification is also essential. Trying to make it more precise is an interesting and challenging enterprise. However, one should not expect a definite answer in the form of a definition here. Rather, I think, the question is a guideline that points out interesting philosophical issues as we shall argue in Section \ref{section:CognitionAndComplexity}. 
\index{Turing machine}
Moreover, these vague quantifiers could be read in other parts of science too. For example, the Second Law of Thermodynamics \index{Second Law of Thermodynamics} tells us that all isolated macroscopic processes in nature are going in the direction that leads to an increase of entropy. First of all, it is per definition not possible to observe perfectly isolated macroscopic processes. So, in all practical applications of the Second Law of Thermodynamics we would have to comfort ourselves with a highly isolated macroscopic process instead. And then, we know, as a matter of fact that we should read a vague quantifier to the effect that \textit{almost all} such processes lead to an increase of entropy. A notable exception is given by processes that involve living organisms. Of course one can debate here to what extent higher-level living organisms can occur in a highly isolated environment. But certainly lower-level living organisms like colonies of protozoans can occur in relative isolation thus at least locally violating the Second Law of Thermodynamics. (See \cite{Bennett} and \cite{WhatIsLife} for different viewpoints on whether life violates the Second Law of Thermodynamics.)

It has been observed before in \cite{NKS} that the $\mathbf{PCE}$ does imply the $\mathbf{CT}$. Note that $\mathbf{PCE}$ quantifies over all processes, be they natural or designed by us. Thus in particular Turing Machines are considered by the $\mathbf{PCE}$ and stipulated to have the maximal degree of computational sophistication which implies the $\mathbf{CT}$ thesis. \index{Church-Turing thesis}\index{Principle of Computational Equivalence}

But the $\mathbf{PCE}$ says more. It says that the space of possible degrees of computational sophistication between obviously simple and universal is practically un-inhibited. In what follows we shall address the question what might cause this. We put forward two observations. First we formulate a natural candidate principle that can account for $\mathbf{PCE}$ and argue for its plausibility. Second, we shall briefly address how cognition can be important. In particular, the way we perceive, interpret and analyze our environment could be such that in a natural way it will not focus on intermediate degrees even if they were there.

We would like to stress here that intermediate degrees refer to undecidable yet not-universal to be on the safe side. There are various natural decidable processes known that fall into different computational classes like P-time and EXP-time processes which are known to be different classes. 

In theoretical computer science there are explicit undecidable intermediate degrees \index{intermediate degrees} known and  the structure of such degrees is actually known to be very rich. However, the processes that generate such degrees are very artificial whence unlikely to be observed in nature. Moreover, although the question about the particular outcome of these processes is known to yield intermediate degrees, various other aspects of these processes exhibit universal complexity.

\subsection{Complexity and Evolution}\index{Darwinian evolution}
In various contexts but in particular in evolutionary processes one employs the principle of Natural Selection, \index{Natural Selection} often also referred to as Survival of the Fittest.\index{survival of the fittest} These days basically everyone is familiar with this principle. It is often described as species being in constant fight with each other over a limited amount of resources. In this fight only those species that outperform others will have access to the limited amount of resources, whence will be able to pass on its reproductive code to next generations causing the selection.

We would like to generalize this principle to the setting of computations. This leads us to what we call the principle of Generalized Natural Selection: \index{Natural Selection, Generalized}

\begin{quote}
{\bf $\mathbf{GNS}$:} In nature, computational processes of high computational sophistication are more likely to maintain/abide than processes of lower computational sophistication.
\end{quote}

If one sustains the view that all natural processes can be viewed as computational ones, this generalization is readily made. For a computation, to be executed, it needs access to the three main resources  space, matter, and time. If now one computation outperforms the other, it will win the battle over access to the limited resources and abide. What does outperform mean in this context? 

Say we have two neighboring processes $\Pi_1$ and $\Pi_2$ that both need resources to be executed. Thus, $\Pi_1$ and $\Pi_2$ will interfere with each other. Stability of a process is thus certainly a requirement for survival. Moreover, if $\Pi_1$ can incorporate, or short-cut $\Pi_2$ it can actually use $\Pi_2$ for its survival. A generalization of incorporating, or short-cutting is given by the notion of simulation that we have given above. Thus, if $\Pi_1$ can simulate $\Pi_2$, it is more likely to survive. In other words, processes that are of higher computational sophistication are likely to outperform and survive processes of lower computational sophistication. In particular, if the process $\Pi_1$ is universal, it can simulate any other process $\Pi_2$ and thus is likely to use or incorporate any such process $\Pi_2$.

Of course this is merely a heuristic argument or an analogy rather than a conclusive argument for the $\mathbf{GNS}$ principle. One can think of experimental evidence where universal automata in the spirit of the Game of Life are run next to and interacting with automata that generate regular or repetitive patterns to see if, indeed, the more complex automata are more stable than the repetitive ones. However one cannot expect of course that experiments and circumstantial evidence can substitute or prove the principle.

Just like the theory of the selfish gene \index{selfish gene} 
(see \cite{SelfishGene}) shifted the scale on which natural selection was to be considered, now $\mathbf{GNS}$ is an even more drastic proposal and natural selection can be perceived to occur already on the lowest possible level: individual small-scale computational processes. 

We note that $\mathbf{GNS}$ only talks about computational processes in nature and not in full generality about computational processes either artificial or natural as was the case in $\mathbf{PCE}$. Under some reasonable circumstances we may see $\mathbf{GNS}$ as a consequence of $\mathbf{PCE}$. For if  $\neg$ $\mathbf{GNS}$ were true, there would be no complex processes to witness after some time and this contradicts $\mathbf{PCE}$. Thus we have:
\[
{\mathbf{PCE}} \ \Longrightarrow \ \mathbf{CT}\  + \ \mathbf{GNS}.
\] 
As we already mentioned, $\mathbf{GNS}$ only involves computational processes in nature. Thus we cannot expect that $\mathbf{CT}+  \mathbf{GNS}$ is actually equivalent to $\mathbf{PCE}$. However, if we restrict $\mathbf{PCE}$ to talk only about processes in nature, let us denote this by $\mathbf{PCE}'$, then we do argue that we can expect a correspondence. That is:
\[
{\mathbf{PCE'}} \ 
\approx
\ \mathbf{CT}\  + \ \mathbf{GNS}.
\] 
But ${\mathbf{PCE'}}$ tells us that almost all computational processes in nature are either simple or universal. If we have ${\mathbf{GNS}}$ we find that more sophisticated processes will outperform simpler ones and ${\mathbf{CT}}$ gives us an attainable maximum. Thus the combination of them would yield that in the limit all processes end up being complex. The question then arises, where do simple processes come from? (Normally, the question is where do complex processes come from, but in the formal setting of $\mathbf{CT}  +\mathbf{GNS}$ it is the simple processes that are in need of further explanation.)

Simple processes in nature often have various symmetries. \index{symmetries} As we have argued above these symmetries are readily broken when a simple system interacts with a more complex one resulting in the simple system being absorbed in the more complex one. We see two main forces that favor simple systems. 

The first driving force is what we may call \textit{cooling down}. For example, temperature/energy going down, or material resources growing scarce. If these resources are not available, the complex computations cannot continue their course, breaking down and resulting in less complex systems.

A second driving force may be referred to as \textit{scaling} and invokes mechanisms like the Central Limit Theorem. The Central Limit Theorem \index{Central Limit Theorem} is a phenomenon that creates symmetry by repeating a process with stochastic outcome a large number of times yielding the well-known Gaussian distribution. Thus the scale (number of repetitions) of the process determines the amount of symmetry that is built up by phenomena that invoke the Central Limit Theorem.

In the above, we have identified a driving force that creates complexity ($\mathbf{GNS}$) and two driving forces that creates simplicity: cooling down and scaling. In the light of these two opposite forces we can restate $\mathbf{PCE'}$ as saying that simplicity and universality are the two main attractors of these interacting forces. 

Note that we deliberately do not speak of an equivalence between $\mathbf{PCE'}$ and $\mathbf{CT}\  + \ \mathbf{GNS}$. Rather we speak of a correspondence. It is like when modeling the movement of a weight on a spring on earth. The main driving forces in this movement are gravitation and the tension of the spring. However, this does not fully determine a final equilibrium if we do not enter in more details taking into account friction and the like. It is in the same spirit that we should interpret the above mentioned correspondence. 

However, there are two issues here that we wish to address. First, we have argued that $\mathbf{CT}\  + \ \mathbf{GNS}$ is in close correspondence to $\mathbf{PCE'}$ which is a weak version of $\mathbf{PCE}$. What can be said about full $\mathbf{PCE}$? In other words, what about those processes that we naturally come up with? There is clearly a strong cognitive component in the study of those processes that we naturally come up with.

Second, $\mathbf{PCE}$ has a strong intrinsic implicit cognitive component as it deals with the processes that we observe in nature and not necessarily the ones that are out there. Admittedly, in its original formulation there is no mention of this cognitive component in $\mathbf{PCE}$ but only the most radical Platonic reading of $\mathbf{PCE}$ would deny that there is an intrinsic cognitive component present.

We shall try to address both issues in the next section where we discuss how cognition enters a treatment of $\mathbf{PCE}$.



\section{Cognition and Complexity}\label{section:CognitionAndComplexity} \index{cognition and complexity}
In the first section we used a robust definition of complexity by saying a process is complex if we can easily perceive it as a universal computational process. In the current section we shall deliberately use a less formal and rigorous definition.

\subsection{Relative complexity}\label{section:CognitionAndComplexity:Relativity}
In the current section we say that a process is complex if comprehending or describing it exceeds or supersedes all available resources (time, space, description size). In doing so the relative nature of complexity becomes apparent. 

The relativity is not so much due to our underspecification when we spoke of comprehension or description of a process. One can easily think of formal interpretations of these words. For example, comprehension can be substituted by obtaining a formal proof in a particular proof system. Likewise, descriptions can be thought of as programs that reproduce or model a process. However for the sake of the current argument it is not necessary to enter into that much detail or formalization.

The relativity of the notion of complexity that we employ in this section is merely reflected in how much resources are available. A problem or process can be complex for one set of resources but easy for another.

For example, let us consider the currently known process/procedure $\Pi$ that decides whether or not a natural number is prime (see \cite{PrimesInP}). If we only have quadratic time resources, then $\Pi(n)$ is difficult as the current known procedure is known to require an amount of time in the order of $|n|^{12}$ (that is, in the order of magnitude of the length of $n$ (written in decimal notation) to the power 12). Of course, if we allow polynomial time, then primality is an easy problem.

This relativity is a rather trivial observation. The point that we wish to make here however is more subtle and profound. So, we depart from the observation that complexity is always relative to the framework in which it is described/perceived. Now, the ultimate framework where all our formal reasoning is embedded is our own framework of cognitive abilities. And this has two important consequences. 

Firstly, it implies that if we study how our cognitive framework deals with complexity and related notions, we get a better understanding of the nature of the definitions of complexity that we come up with. And secondly, it strongly suggests that various notions and definitions of complexity in various unrelated areas of science in the end are of the same nature. Thus, various formal theorems that relate different notions of complexity, like ergodicity, entropy, Kolmogorov-Chaitin complexity, computational complexity etc. are expected to be found. And as a matter of fact, in recent developments many such theorems have been proven. In the final section of this chapter we shall see a new such and rather unexpected connection between two seemingly different notions of complexity: fractal dimensions versus computational runtime classes.

\subsection{Cognitive diagonalization}

In this section we wish to dwell a bit on the following simple observation: as human beings we have a natural ability to consider a system in a more complete and more complex framework if necessary. We shall give some examples of this in a formalized setting and pose the question how we naturally generate more complex systems in a cognitive and less formal setting.

Suppose we study a system $\mathcal{S}$ with just an initial element --let us call that 0-- and an operator $S$ that gives us a next, new and unique element which we call the successor. The smallest system of this kind can be conceived as the natural numbers without further extra structure on them:
\[
\{  0, S0, SS0, SSS0, \ldots  \}.
\] 
If we study this system in a fairly simple language
it turns out that all questions about this systems are easily decidable by us.

Of course, we would not leave it there and move on to summarize certain processes in this system. The process of repeating taking the successor a number of times is readily introduced yielding our notion of addition defined by $x + 0 =x $ and $x+Sy = S(x+y)$. So, by summarizing certain processes in $\mathcal{S}$ we naturally arrive at a richer structure $\mathcal{S}'$ whose operations are $S$ and $+$. 

If we now study $\mathcal{S}'$ in the same simple language as we used for $\mathcal{S}$ but now with the additional symbol $+$, we see that all questions are still decidable. That is, we can still find the answer to any question we pose about this system in an algorithmic way. The time complexity of the algorithm is slightly higher than that of $\mathcal{S}$ but the important thing is that it is still decidable.

By a process similar to by which we went from $\mathcal{S}$ to $\mathcal{S}'$ we can now further enrich our structure. We consider repeated addition to get to our well-known definition of multiplication: $x \times 0 = 0$ and $x\times Sy = x\times y + x$. The resulting structure $\mathcal{S}''$ has now three operations: $S, +$ and $\times$. However questions about this structure in this language now turn out to be \textit{undecidable}. That is, there is no single algorithm that settles all questions about this structure in our simple language. \index{undecidability}

The process by which we went from $\mathcal{S}$ to the more complex system $\mathcal{S}'$ and from $\mathcal{S}'$ to the more complex system $\mathcal{S}''$ is called iteration. One may think that this can only be repeated $\omega$ many times but we shall now describe a more general process of gaining complexity which is called \textit{diagonalization} \index{diagonalization} and of which iteration is just a special case. As an illustration of how this works we shall give a proof of G\"odel's First Incompleteness Theorem. \index{G\"odel's First Incompleteness Theorem} 

\begin{quote}
{\bf G\"odel's First Incompleteness Theorem}\index{G\"odel's First Incompleteness Theorem} For each algorithmically enumerable theory $T$ that only proves true statements and that is of some minimal strength there is a true sentence $\varphi_T$ that is not provable in $T$.
\end{quote}

Although G\"odel had a slightly different formulation of his First Incompleteness Theorem in essence it is the one that we shall prove here. Our proof will focus on the computable functions $f(x)$ that $T$ can prove to be total. Thus, we focus on those unary functions $f$ which are computable and moreover, so that $T$ proves that $f$ is defined for each value of $x$. We shall write  
\[
T\vdash \forall x \, \exists y \ f(x)=y
\]
for the latter. As $T$ is algorithmically enumerable, we can fix an enumeration and just enumerate the proofs of $T$ and stick with all the proofs $\pi_i$ that prove some computable function $f_i$ to be total. Once we have a way to make this list of functions $f_i$ we readily come up with a new computable function $f'$ which is total but not provably so by $T$. We construct $f'$ by what is called diagonalization \index{diagonalization} and it will soon become clear why this is called this way. We can make a table of our functions $f_i$ with their values where we in anticipation have displayed the diagonal in boldface.

\begin{table}[htdp]
\begin{center}
\begin{tabular}{|c|c|c|c| c }
\hline
$\bf f_0(0)$ & $f_0(1)$ & $f_0(2)$ & $f_0(3)$ & \ldots \\
\hline
$f_1(0)$ & $\bf f_1(1)$ & $f_1(2)$ & $f_1(3)$ & \ldots \\
\hline
$f_2(0)$ & $f_2(1)$ & $\bf f_2(2)$ & $f_2(3)$ & \ldots \\
\hline
$f_3(0)$ & $f_3(1)$ & $f_3(2)$ & $\bf f_3(3)$ & \ldots \\
\hline
$\vdots$ & $\vdots$&$\vdots$ &$\vdots$ & $\ddots$
\end{tabular}
\end{center}
\label{default}
\end{table}%

We now define $f'(x) = f_x(x)+42$. Clearly $f'$ differs from any $f_i$ as it differs on the diagonal. However, $f'$ is clearly a total function and there is also an easy algorithm to compute it: to compute $f(x)$ we enumerate, using the fixed enumeration of the theorems of $T$, all proofs of $T$ until we arrive at $\pi_x$. Then we compute $f_x(x)$ and add 42 to the result. 

To summarize, we have provided a total computable function that is not proven to be total by $T$ whence $T$ is incomplete. It is clear what minimal requirements should be satisfied by $T$ in order to have the proof go through.

For the main argument of this paper this proof of G\"odel's First Incompleteness Theorem is not entirely necessary. We have included it for two main reasons. Firstly, of course, there is the beauty of the argument which is a reason for itself. And second, the proof illustrates nicely how diagonalization works.

In mathematical logic diagonalization is a widely used technique and a universal way to obtain more complex systems. An interesting and important question is, is there a natural cognitive counterpart of this? So, is there some sort of universal cognitive construct --cognitive diagonalization \index{cognitive diagonalization} if it were-- that always yields us a more complex framework in which to study a system. For it is clear that we tend to add complexity to systems that we build and perceive until it reaches the boundaries of our abilities. And thus we pose the question if there is some universal and natural way by which we add this complexity. 

As an academic exercise one could try to rephrase diagonalization in a setting of formalized cognition but that would yield a very artificial principle. Moreover this will say nothing about what we actually do in our heads.

\subsection{Cognition, complexity and evolution}

Fodor \index{Fodor, J. A.} has postulated a principle concerning our language. It says that (see \cite{FrideonsFodor}) the structure and vocabulary of our language is such that it is efficient in describing our world and dealing with the frame problem. \index{frame problem} The frame problem is an important problem in artificial intelligence which deals with the problem how to describe the world in an efficient way so that after a change in the state of affairs no entirely new description of the world is needed.\index{cognition}

On a similar page, we would like to suggest that our cognitive toolkit has evolved over the course of time so that it best deals with the processes it needs to deal with. Now, by $\mathbf{PCE}$ these processes are either universal or very simple. Thus, it seems to make sense in terms of evolution to have a cognitive toolkit that is well-suited to deal with just two kinds of processes: the very simple ones and the universal ones.

Thus, it could well be that there actually are computational processes out there that violate $\mathbf{PCE}$ just as there are chemical processes (life) that locally violate the Second Law of Thermodynamics but that our cognitive toolkit is just not well-equipped enough to deal with them.

This might also be related to the question we posed in the previous section: how do we add complexity to a system? Let us continue the analogue with formal logic. Diagonalization is currently the main universal tool for adding strength to a system. However, there are various indications that for many purposes diagonalization seems not to be fine-grained enough and some scientists believe this is one of the main reasons why we have such problems dealing with the famous $\mathbf{P}$ versus $\mathbf{NP}$ problem. Likewise, it might be that cognitive diagonalization is not fine-grained enough to naturally observe/design intermediate degrees.


\section{Complexity everywhere: small Turing machines}\label{section:TMs}

In this final section I will report on an ongoing project jointly with Fernando Soler-Toscano\index{Soler-Toscano, F.} and Hector Zenil\index{Zenil, H.}. In this project we study the structures that arise when one considers small Turing machines. Here, in this final section we relax the working definition of complexity even further to just refer to interesting structures.

In 2009 I attended the NKS summer school led by Stephan Wolfram in Pisa, Italy. One of the main themes of NKS is that simple programs can yield interesting and complex behavior. Being trained as a mathematician and logician this did not at all shock my world view as there are various simple functions or axiomatic systems known that yield very rich and complex structures. However, when you start delving the computational universe yourself it is that you get really excited about the NKS paradigm. It is not merely that there are various interesting systems out there, it is the astonishing fact that these systems abound. And wherever you go and look in the computable universe you find beautiful, intriguing and interesting structures. In this final section I shall report on one of those explorations in the computational universe.

The set-up of our experiment was inspired by an exploration performed in \cite{NKS} and we decided to look at small Turing-machines. There are various definitions of Turing machines \index{Turing machine} in the literature which all look alike. For us, a Turing machine (TM) consist of a tape of cells where the tape extends infinitely to the left and is bounded to the right. Each cell on the tape can be either black (1) or white (0) and this start configuration is specified by us. There is a \textit{head} that moves over the tape and as it does so, the head can be in one of finitely many \textit{states} (like states of mind). 

We have now specified the hardware of a TM. The software, so to say, of a TM consists of a lookup table. This table tells the head what to do in which situation. More concrete, depending on the state the head is in and depending what symbol the head reads on the cell of the tape it is currently visiting, it will perform an action as specified by the lookup table. This action is very simple and consist of three parts: writing a symbol on the cell it currently is at, moving the head one cell left or right and going to some state of mind.

We only looked at small Turing machines that have either 2, 3 or 4 states of mind. On those machines we defined a computation to start at the right-most cell of the tape in State 0. We say the computation halts when the head `drops off' at the right-hand side of the tape. That is, when it is at the border cell of the tape and receives a command to go one cell to the right. We fed these TMs successive inputs that were coded in unary plus one. Thus, input 0 was coded by just one black cell, input 1 was coded by two consecutive black cells, and input $n$ was coded by $n+1$ consecutive black cells on an otherwise white tape.

With this set-up we looked at the different functions that were computed by these small TMs and had a particular focus on the runtimes that occurred. Of course, there are various fundamental issues to address that are mostly related to either the Halting Problem \index{Halting Problem} (there is no algorithm that decides whether a TM will halt on a certain input) or unfeasibility. Some of these issues are addressed in \cite{ZJS2010}.

\begin{figure}[htb!]
  \centering
  \includegraphics[width=12.1cm]{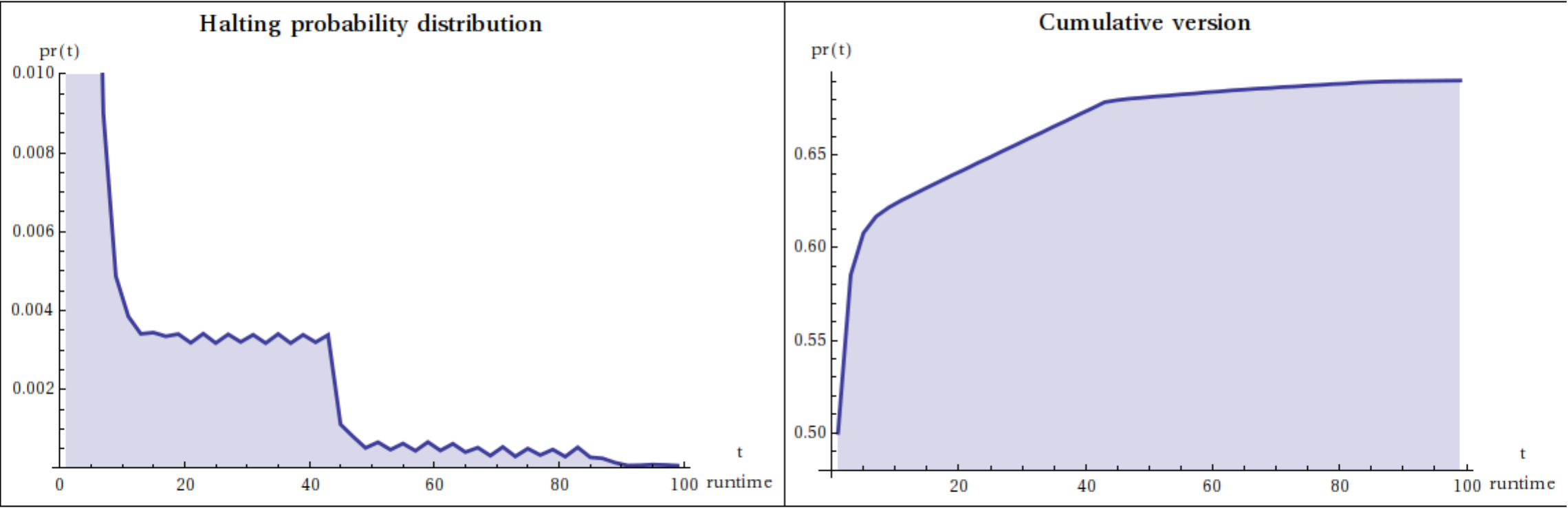}
\caption{Halting time distribution among TMs with three states and two colors on the first 21 inputs.}
  \label{fig:occrun32}
\end{figure}

When plotting the halting probability distribution for our TMs we verified a theoretical result by Calude \index{Calude, C.} to the effect that most TMs either halt quickly or they never halt at all (\cite{calude}). Although this result was expected we did not expect the pronounced phase-transitions \index{phase transitions} one can see in Figure \ref{fig:occrun32} in the halting probability distributions that we found. In a sense, these phase transitions are rudimentary manifestations of the low-level complexity classes as described in \cite{JoostenSolerZenil2011}.

Another striking feature that we found is that TMs tend to grow \textit{slower} if you give them more resources. Let us make this statement more precise. We studied the behavior of all 4,096 TMs with two colors and two states (we speak of the (2,2)-space). In total, they computed 74 different functions. We also studied the behavior of all the 2,985,984 TMs with two colors and three states where now 3,886 different functions were computed. Any function that is computed in the (2,2)-space is easily seen to be also present in (3,2)-space. We looked at the time needed to compute a function in the different spaces. To our surprise we saw that almost always slow-down occurs. And at all possible levels: slow-down on average, worst case, harmonic average, asymptotically, etc. We only found very few cases of at most linear speed-up.

\begin{figure}[htb!]
  \centering
  \includegraphics[width=5cm]{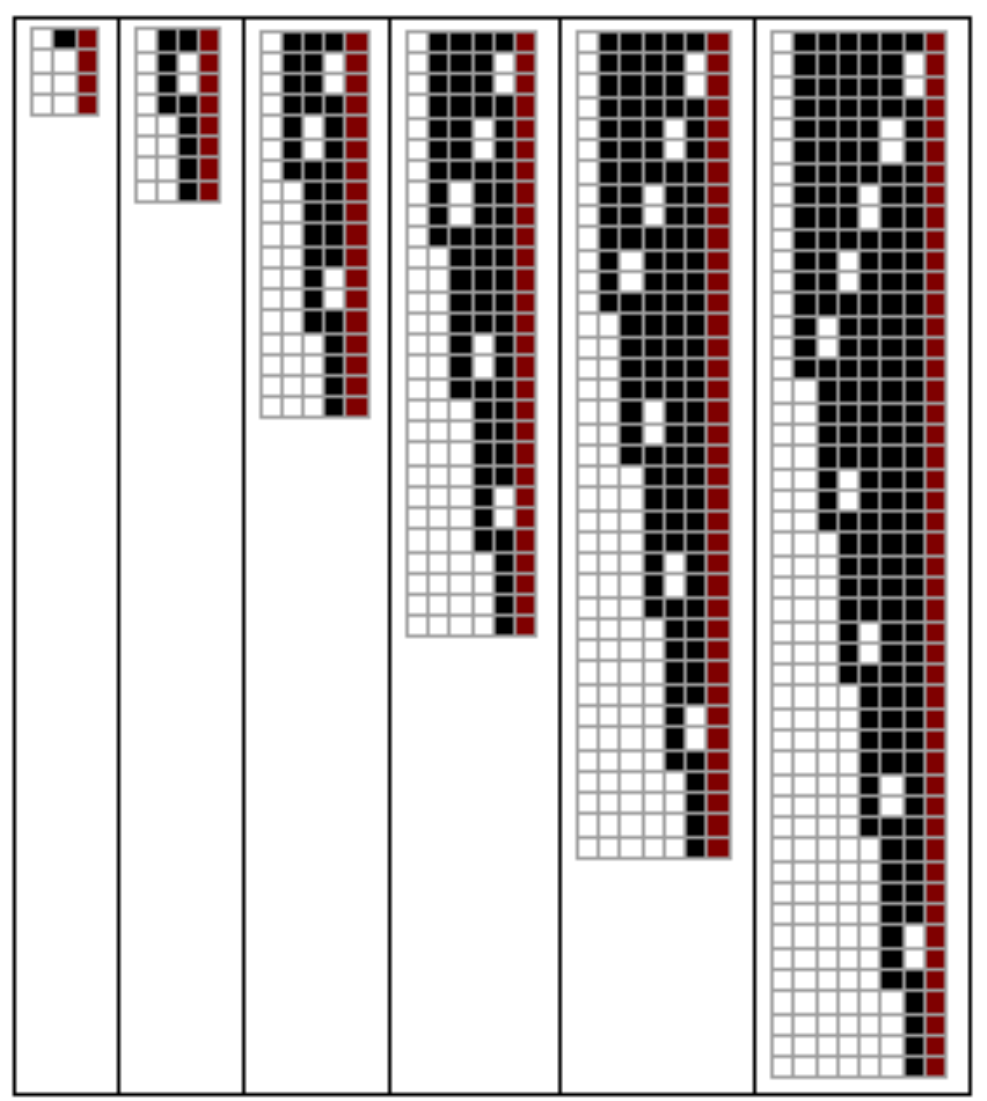}
  \includegraphics[width=3.2cm]{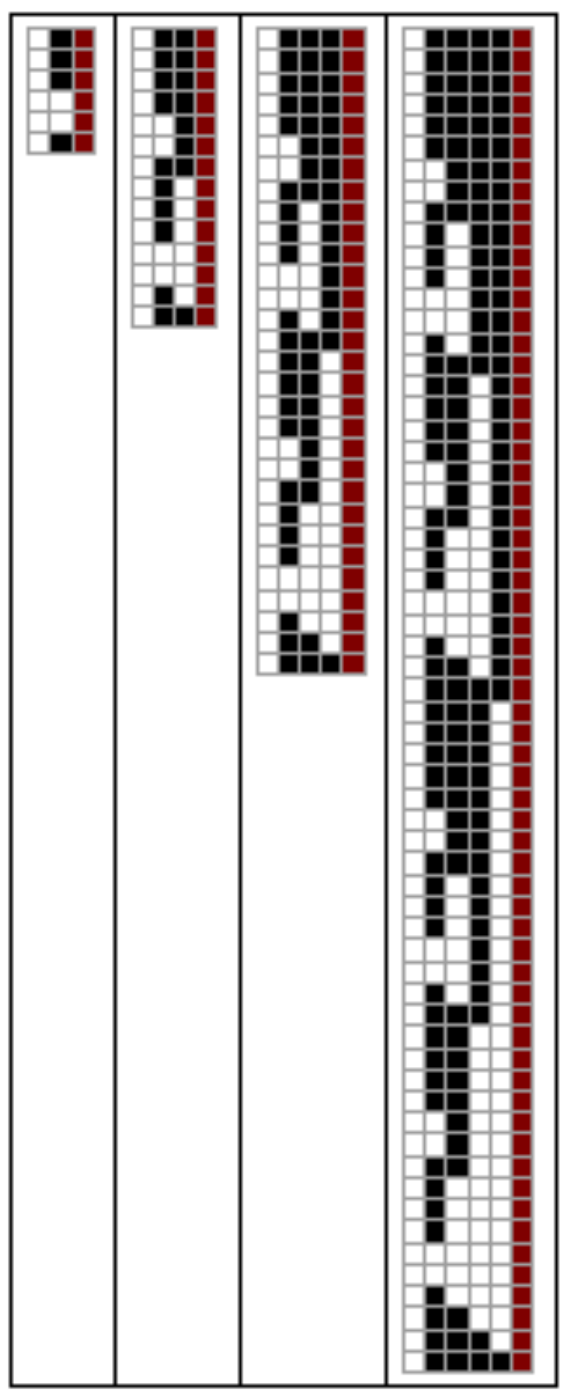}
  \caption{Tape evolution for Rules 2205 (left) and 1351 (right).}
  \label{fig:tape2205and1351}
\end{figure}

So the overall behavior of these small TMs revealed interesting structures to us. But also looking at each particular TM showed interesting structures. In Figure \ref{fig:tape2205and1351} we show two such examples. The rule numbering refers to Wolfram's enumeration scheme for (2,2) space as explained in \cite{NKS} and \cite{JoostenDemonstration2010}.

For TM number 2205 we have plotted the tape evolution for the first 6 entries. So, each gridded rectangle represents a complete computation for a certain input. The diagrams should be interpreted as follows. The top row represents the initial tape configuration. The white cells represent a zero and the black cells a one. The grey cell represent the edge of the tape. Now each row in the gridded rectangle depicts the tape configuration after one more step in the computation. That is why each row differs at at most one cell from the previous row. We call these rectangles space-time diagrams of our computation where the space/tape is depicted horizontally and the time vertically. 

We now see that TM 2205 always outputs just one black cell. Its computation yields a space-time diagram with a very clear localized character where the head has just moved from right to the left-end of the input and back to the right end again doing some easy computation in between. TM number 1351 shows a clear recursive structure. Curiously enough this machine computes a very easy function which is just the tape identity. So it does a dazing amount of things (it needs exponential time for it) to leave in the end (the bottom row) the tape in the exact same configuration as the input (the top row). For more examples and structure we refer the interested reader to \cite{JoostenSolerZenil2011}.

Let us take a closer look at our pictures from Figure \ref{fig:tape2205and1351}. It is clear that each TM defines an infinite sequence of these space-time diagrams as each different input defines such a diagram. It is pretty standard to assign to this sequence of space-time diagrams a fractal dimension $d_{\tau}$ that describes some features of the asymptotic behavior of a TM $\tau$. We have empirically established for all TMs in (2,2)-space a very  curious correspondence. It turns out that 
\begin{quote}
The fractal dimension $d_{\tau}$ that corresponds to a TM $\tau$ is 2 if and only if $\tau$ computes in linear time. The dimension $d_{\tau}$ is 1 if and only if $\tau$ computes in exponential time. 
\end{quote}
This result is remarkable because it relates two completely different complexity measures: the geometrical fractal dimension on the one side versus the time complexity of a computation on the other side. The result is one out of the many recent results that link various notions of complexity the existence of which we already forecasted on philosophical grounds in Section \ref{section:CognitionAndComplexity:Relativity}.

\section*{Acknowledgments}
The author would like to thank David Fern\'andez-Duque, Jos\'e Mart\'{\i}nez-Fern\'andez, Todd Rowland, Stephen Wolfram, and Hector Zenil for their comments, discussions and general scientific input and feedback for this paper.


\begin{thebibliography}{99}

\bibitem{PrimesInP}
M. Agrawal, N. Kayal, N. Saxena. ``PRIMES is in P". \textit{Annals of Mathematics} 160 (2): 781Ð793, 2004.

\bibitem{Bennett} C. H. Bennett. The Thermodynamics of Computation - A Review. \textit{Int. J. Theoretical Physics} vol. 21, no. 12, pp. 905-940, 1982.


\bibitem{calude}  C.S. Calude, M.A. Stay, Most programs stop quickly or never halt, \textit{Advances in Applied Mathematics}, 40, p 295-308, 2005.

\bibitem{MembraneComputing} C. S. Calude, and G. Paun. \textit{Computing with Cells and Atoms: An Introduction to Quantum, DNA and Membrane Computing}, CRC Press, 2000.


\bibitem{SelfishGene} R. Dawkins. \textit{The Selfish Gene}, New York City: Oxford University Press. ISBN 0-19-286092-5, 1976.

\bibitem{FrideonsFodor}
J. A. Fodor. Modules, Frames, Fridgeons, Sleeping Dogs, and the Music of the Spheres, in \textit{Pylyshyn}, 1987.

\bibitem{JoostenDemonstration2010} J. J. Joosten. Turing Machine Enumeration: NKS versus Lexicographical, \textit{Wolfram Demonstrations Project}; \texttt{http://demonstrations.wolfram.com/\\TuringMachineEnumerationNKSVersusLexicographical/}, 2010.

\bibitem{JoostenSolerZenil2011}
J. J. Joosten, F. Soler, and H. Zenil. Program-size versus Time Complexity. Slowdown and Speed-up Phenomena in the Micro-cosmos of Small Turing Machines. \textit{Int. Journ. of Unconventional Computing}, Vol. 7, pp. 353-387, 2011.

\bibitem{JoostenSolerZenilDemonstration} J. J. Joosten, F. Soler Toscano, H. Zenil. Turing machine runtimes per number of states, to appear in \textit{Wolfram Demonstrations Project}, 2012.

\bibitem{QuantumComputing}Michael A. Nielsen, and Isaac L. Chuang.
\textit{Quantum Computation and Quantum Information}, Cambridge University Press, 2000.


\bibitem{DNAcomputing} G. Paun, G. Rozenberg, and A. Salomaa. \textit{DNA Computing: New Computing Paradigms}, Springer, 2010.


\bibitem{Hypertasks} I. Pitowsky. The Physical Church Thesis and Physical Computational Complexity. \textit{Iyyun, A Jerusalem Philosophical Quarterly}, 39, 81-99, 1990.

\bibitem{WhatIsLife} E. Schr\"odinger. \textit{What is Life?} Cambridge University Press, 1944.


\bibitem{NKS} S. Wolfram, \textit{A New Kind of Science}, Wolfram Media, 2002.

\bibitem{ZJS2010} H. Zenil, F. Soler Toscano and, J. J. Joosten. Empirical encounters with computational irreducibility and unpredictability, \textit{Minds and Machines}, vol. 21, 2011.
\end{thebibliography}
\end{document}